\documentclass[sigconf]{acmart}

\renewcommand\footnotetextcopyrightpermission[1]{} 
\usepackage{booktabs} 
\usepackage{rotating} 
\usepackage{cleveref} 

\usepackage{xcolor}
\usepackage{pgfplots}
\usepackage{tikz}
\usepackage[colorinlistoftodos,prependcaption]{todonotes}
\let\svtodo\todo\renewcommand\todo[1]{\svtodo[inline]{#1}}
\newcommand{\forward}[1]{\overrightarrow{#1}}
\newcommand{\backward}[1]{\overleftarrow{#1}}
\newcommand{\token}[1]{<#1>}
\DeclareMathOperator*{\argmax}{arg\,max}

\setcopyright{none}

\begin{document}
\title{Generating Commit Messages from Git Diffs}

\author{Sven van Hal}
\affiliation{%
 \institution{Delft University of Technology}}
 \email{S.R.P.vanHal@student.tudelft.nl}
 
 \author{Mathieu Post}
\affiliation{%
 \institution{Delft University of Technology}}
 \email{M.Post@student.tudelft.nl}

\author{Kasper Wendel}
\affiliation{%
 \institution{Delft University of Technology}}
 \email{K.Wendel@student.tudelft.nl}

\begin{abstract}
Commit messages aid developers in their understanding of a continuously evolving codebase. However, developers not always document code changes properly. Automatically generating commit messages would relieve this burden on developers.

Recently, a number of different works have demonstrated the feasibility of using methods from neural machine translation to generate commit messages. This work aims to reproduce a prominent research paper in this field, as well as attempt to improve upon their results by proposing a novel preprocessing technique.

A reproduction of the reference neural machine translation model was able to achieve slightly better results on the same dataset. When applying more rigorous preprocessing, however, the performance dropped significantly. This demonstrates the inherent shortcoming of current commit message generation models, which perform well by memorizing certain constructs.

Future research directions might include improving diff embeddings and focusing on specific groups of commits.
\end{abstract}


\acmConference[Machine Learning and Analytics for Software Engineering]{Machine Learning and Analytics for Software Engineering}{October 2019}

\settopmatter{printacmref=false}
\pagestyle{plain}
\maketitle

\section{Introduction}
Software development is a continuous process: developers incrementally add, change or remove code and store software revisions in a Version Control System (VCS). Each changeset (or: \textit{diff}) is provided with a commit message, which is a short, human-readable summary of the \textit{what} and \textit{why} of the change \citep{buse_automatically_2010}.

Commit messages document the development process and aid developers in their understanding of the state of the software and its evolution over time. However, developers do not always properly document code changes or write commit messages of low quality \citep{dyer_boa:_2013}. This applies to code documentation in general and negatively impacts developer performance \citep{roehm_how_2012}. Automatically generating high-quality commit messages from diffs would relieve the burden of writing commit messages off developers, and improve the documentation of the codebase.

As demonstrated by \citet{buse_automatically_2010}, \citet{cortes-coy_automatically_2014} and \citet{shen_automatic_2016}, predefined rules and text templates can be used to generate commit messages that are largely preferred to developer-written messages in terms of descriptiveness. However, the main drawbacks of these methods are that 1) only the \textit{what} of a change is described, 2) the generated messages are too comprehensive to replace short commit messages, and 3) these methods do neither scale nor generalize on unseen constructions, because of the use of hand-crafted templates and rules.

According to a recent study by \citet{jiang_towards_2017}, code changes and commit messages exhibit distinct patterns that can be exploited by machine learning. The hypothesis is that methods based on machine learning, given enough training data, are able to extract more contextual information and latent factors about the \textit{why} of a change. Furthermore, \citet{allamanis_survey_2018} state that source code is ``a form of human communication [and] has similar statistical properties to natural language corpora''. Following the success of (deep) machine learning in the field of natural language processing, neural networks seem promising for automated commit message generation as well.

\citet{jiang_automatically_2017} have demonstrated that generating commit messages with neural networks is feasible. This work aims to reproduce the results from \cite{jiang_automatically_2017} on the same and a different dataset. Additionally, efforts are made to improve upon these results by applying a different data processing technique. More specific, the following research questions will be answered:

\begin{itemize}
    \item[RQ1:] \textbf{Can the results from \citet{jiang_automatically_2017} be reproduced?}
    \item[RQ2:] \textbf{Does a more rigorous dataset processing technique improve the results of the neural model?}
\end{itemize}

This paper is structured as follows. In \Cref{sec:background}, background information about deep neural networks and neural machine translation is covered. In \Cref{sec:relatedwork}, the state-of-the-art in commit message generation from source code changes is reviewed. \Cref{sec:methodology} describes the implementation of the neural model. \Cref{sec:preprocessing} covers preprocessing techniques and analyzes the resulting dataset characteristics. \Cref{sec:results} presents the evaluation results and \Cref{sec:discussion} discusses the performance and limitations. \Cref{sec:conclusion} summarizes the findings and points to promising future research directions.

\section{Background}\label{sec:background}
\subsection{Neural Machine Translation}
A recent development in deep learning is sequence-to-sequence learning for Neural Machine Translation \cite{sutskever2014sequence, cho2014learning}. Translation can be seen as a probabilistic process, where the goal is to find a target sentence $\mathbf{y} = (y_1,...,y_n)$ from a source sequence $\mathbf{x} = (x_1,...,x_m)$ that maximizes the conditional probability of $\mathbf{y}$ given the source sentence $\mathbf{x}$, mathematically depicted as $\argmax_y p(y|x)$ \cite{bahdanau2014neural}. But this conditional distribution is of course not given and has to be learned by a model from the supplied data.  
\citeauthor{sutskever2014sequence} \cite{sutskever2014sequence} and \citeauthor{cho2014learning} \cite{cho2014learning} both have proposed a structure to learn this distribution: a model that consists of a encoder and decoder component that are trained simultaneously. The encoder component tries to encode the variable length input sequence $\mathbf{x}$ to a fixed length hidden state vector. This hidden state is then supplied to the decoder component that tries to decode it into the variable length target sequence $\mathbf{y}$. A Recurrent Neural Network (RNN) is used in this to sequentially read the variable length sequence and produces a fixed size vector.

Over the years, different architectural changes of encoder and decoder components were proposed. \citeauthor{sutskever2014sequence} \cite{sutskever2014sequence} introduces multi-layered RNN with Long-Short-Term memory units (LSTM), where both the encoder and decoder consist of multiple layers (4 in their research). Each layer in the encoder produces a fixed size hidden state that is passed onto the corresponding layer in the decoder, where the results are combined into a target sequence prediction. One unexplainable factor noted by the authors of this architecture is that it produces better results if the source sequence is reversed. Note that in each step during decoding, the LSTM only has access to the hidden state from the previous timestep, and the previous predicted token.

\citeauthor{cho2014learning} \cite{cho2014learning} uses a slightly different approach in their model and uses Gated Recurrent Units (GRU) as RNN components. The encoder reads the source sequence sequentially and produces a hidden state, denoted as the context vector. Decoding of a token can now be done based on the previously hidden state of the decoder, the previous predicted token, and the generated context vector. The intuition of this architecture is that it reduces information compression as each decoder step has access to the whole source sequence. The decoder hidden states now only need to retrain information that was previously predicted.

Still, the performance of this process suffers when input sentences start to increase and the information can not be compressed into hidden states \cite{cho2014learning}. \citeauthor{bahdanau2014neural} \cite{bahdanau2014neural} therefore extended the encoder decoder model such that it learns to align and translate jointly with the help of attention. At each time decoding step, the model searches for a set of position in the source sentence where the most relevant information is concentrated. The context vectors corresponding to these positions and the previous generated predicted tokens are then used for prediction of the target sequence. It is also possible to compute attention in different ways as shown by \citeauthor{luong2015effective} \cite{luong2015effective}.

\subsection{Evaluation Metrics}
BLEU \citep{papineni_bleu:_2001} is the most frequently used similarity metric to evaluate the quality of machine translations. BLEU measures how many word sequences from the reference text occur in the generated text and uses a (slightly modified) \textit{n-gram precision} to generate a score. Sentences with the most overlapping n-grams score highest. BLEU can be used to calculate the quality of an entire set of <reference,generated> text pairs, which enables researchers to accurately compare the performance of different models on the same dataset. BLEU can be configured with different n-gram sizes, which is denoted by BLEU-\textit{n} (e.g. BLEU-4).

Another widely used metric is ROUGE \citep{lin_rouge:_2004}. ROUGE can be used to calculate recall and F1 scores in addition to precision. This is done by looking at which n-grams in the generated text occur in the reference text. ROUGE is often used to evaluate the quality of machine-generated text summaries, where a word-for-word reproduction of the reference text that gives a high BLEU score is not appreciated. Still, the generated summary should reflect the original text. ROUGE has a number of specialized extensions, of which ROUGE-L is most appropriate to evaluate commit messages. ROUGE-L measures the longest common subsequence between messages to ``capture the sentence level structure in a natural way'' \citep{lin_rouge:_2004}.

Lastly, METEOR is a similarity metric that uses the harmonic mean between precision and recall. METEOR attempts to correct a number of issues with BLEU, such as the fact that sentences have to be identical to get the highest score and that a higher BLEU score not always equals a better translation. The metric is computed using ``a combination of unigram-precision, unigram-recall, and a measure of fragmentation that is designed to directly capture how well-ordered the matched words in the machine translation are in relation to the reference'' \citep{banerjee_meteor:_2005}.

\section{Related Work}\label{sec:relatedwork}

The first works about commit message generation were published independently at the same time by \citet{loyola_neural_2017} and \citet{jiang_automatically_2017}. Both approaches feature a similar attentional RNN encoder-decoder architecture.

\citet{loyola_neural_2017} use a vanilla encoder-decoder architecture, similar to the architecture \citet{iyer_summarizing_2016} used for code summarization. The encoder network is simply a lookup table for the input token embedding. The decoder network is a RNN with dropout-regularized \textit{long short-term memory} (LSTM) cells. Dropout is also used at the encoder layer and reduces the risk of overfitting on the training data. A global attention model is used to help the decoder focus on the most important parts of the diffs.

\citet{jiang_automatically_2017} propose a more intricate architecture, where the encoder network is also a RNN. This way, the token embedding can be trained for better model performance. The authors do not implement the network themselves, but instead use \textsc{Nematus}, a specialized toolkit for neural machine translation \cite{sennrich_nematus:_2017}. Besides using dropout in all layers, Nematus also uses the computationally more efficient GRU cells instead of LSTM cells.

\citet{liu_neural-machine-translation-based_2018} investigate the model and results of \cite{jiang_automatically_2017} and found that memorization is the largest contributor to their good results. For almost all correctly generated commit messages, a very similar commits was found in the training set. By removing the noisy commits, the model performance drops by 55\%. To illustrate the shortcomings, \citet{liu_neural-machine-translation-based_2018} propose \textit{NNGen}, a naive nearest-neighbor based approach that re-uses commit messages from similar diffs. \textit{NNGen} outperforms \cite{jiang_automatically_2017} by 20\% in terms of BLEU score, which underlines the similarity in the training and test sets.

The most recent work on commit message generation, by \citet{liu_generating_2019}, states that the main drawback of the earlier approaches by \citet{loyola_neural_2017} and \citet{jiang_automatically_2017} is the inability to generate out-of-vocabulary (OOV) words. Commit messages often contain references to specific class names or functions, related to unique code from a project. When this identifier is omitted from a predicted commit message, it might not make sense.

To mitigate this problem, pointer-generator network \textsc{PtrGNCMsg} is introduced: an improved sequence-to-sequence model that is able to copy words from the input sequence with a certain probability \cite{liu_generating_2019}. The network uses an adapted attentional RNN encoder-decoder architecture, where, at each decoder prediction step, ``the RNN decoder calculates the probability $p_\text{copy}$ of copying words from the source sentence according to the attention distribution, and the probability $p_\text{vocab}$ of selecting words in the vocabulary'' \cite{liu_generating_2019}. By combining the vocabulary, input sequence and probabilities, the model is able to generate valid commit messages containing OOV words.

\section{Approach}\label{sec:methodology}

\subsection{Data Collection}
A dataset was created for the Java and the C\# programming language. To gather these, the Github API was used to retrieve the top 1000 most-starred repositories for both languages. The choice to separate these datasets is due to the fact that a sequence to sequence deep learning model was trained, as will be explained in \Cref{subsec:model}. The tokens in the input sequence, the git diff files, must originate from the same distribution of input data. Therefore, the decision was made to train a model per programming language. Also, commit messages can be structured differently per programming language as coding convention vary. Therefore, training a model per language could lead to more accurate git commit message generation.


For each repository, the most recent commits of the default branch (the \textit{master} branch in most cases) were retrieved. For each commit, the commit message and the raw \textit{git diff} output of the commit and it's parent commit were saved. Commits that did not fulfill the following criteria were ignored:
\begin{itemize}
    \item With more than one parent commit (merge commits).
    \item Without a parent commit (initial commits).
    \item With diffs bigger than 1MB. 
\end{itemize}
Also, all messages and diff were encoded in \textit{UTF-8}. Characters not supported by this encoding are replaced with the unicode replacement character. All commits in the default branch, beginning with the most recent one, were collected until all commits in the branch were considered or until 10K commits were collected for that repository.
This results in a dataset for Java with 610.484 messages with diffs and a dataset for C\# with 1.572.274 messages with diffs.

\subsubsection{Training, Validation and Test Set Splits}
The collected data needs to be divided into three distinct sets that have no overlap. These sets will be used as training, validation or testing data respectively. Also, the collected dataset is rather large in size compared to the dataset that \citeauthor{jiang_automatically_2017} used. Therefore, the decision was made to select a equal amount, 36000 objects, at random from the collected dataset. The splits were then created from this subset of 36000 objects according to a ratio of $0.8$, $0.1$ and $0.1$ for training, validation and testing respectively. This division applies for all dataset that were either collected or preprocessed according to our procedure.

\subsection{Encoder-Decoder Model}\label{subsec:model}
\begin{figure}
    \centering
    \includegraphics[width=0.5\textwidth]{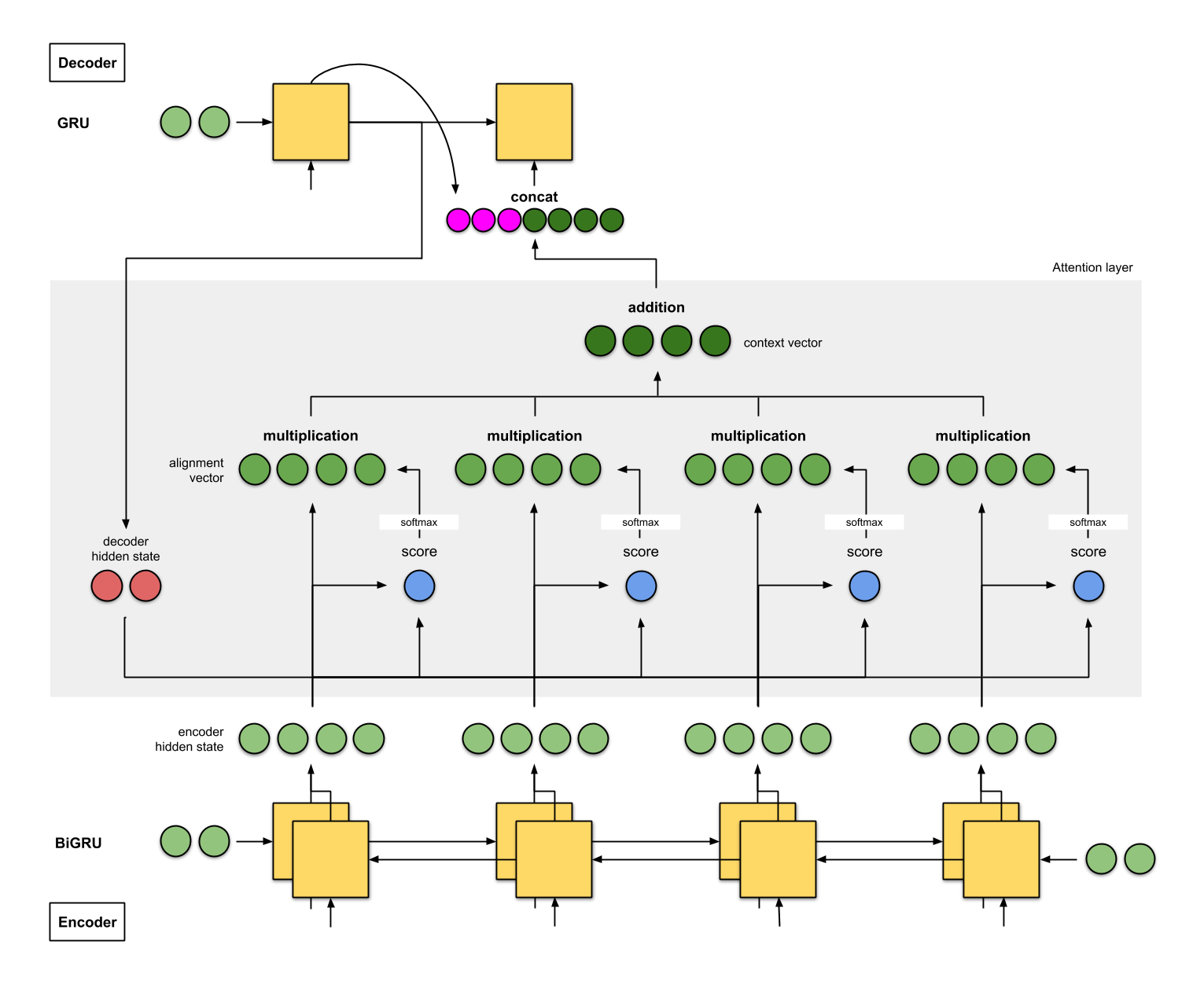}
    \caption{The encoder-decoder model that is used for this research. Both the encoder and decoder consist of a GRU, and the encoder is bidirectional. At each time step, a score is computed for each encoder hidden state based on the previous decoder hidden state. This score is changed to a probability (softmax) and multiplied with the corresponding hidden state to yield a context/alignment vector. The input of the decoder for the nex time step is then the sum of these context vectors  and the previous output of the decoder. Image from \cite{network_pic}.}
    \label{fig:model}
\end{figure}

The approach to generate commit message from a \textsc{git diff} file will be the same as \citeauthor{jiang_automatically_2017}, namely with the neural machine translation approach that does sequence to sequence learning. The model that is to be used is from \citeauthor{bahdanau2014neural}, a encoder-decoder model that uses attention to attend to the important parts of the \textit{git diff} sequence $\mathbf{x}$ during the generation of the commit message sequence $\mathbf{y}$. A schematic view is given in \Cref{fig:model}. Each of the components of the model will be discussed more thoroughly in the sections below.

\subsubsection{Encoder}
The encoder processes the variable length source sequence $x = (x_1,...,x_t)$ one token at a time until it reaches the end of the sequence. At each time step $t$, the hidden state $h_t$ is computed as:

$$h_t = f(h_{t-1},x_t)$$

The function $f$ is a non-linear function such as a LSTM or GRU, which both have the same characteristic of providing memory. In this research, the bidirectional GRU was used according to the network of \cite{bahdanau2014neural} that encode the sequence in both a forward manner as $\forward{h_t}$, and backward as $\backward{h_t}$. This means the hidden state at each time step $t$ consists of a concatenation of both states: $h_t= [\forward{h_t},\backward{h_t}$]. Note that each sequence contains a start (\token{sos}) and stop (\token{eos}) symbol, and thus $\forward{x_0} = \token{sos}$ and $\backward{x_0} = \token{eos}$ are the initial input tokens. If all tokens of the sequence are seen by the encoder then the hidden states are passed to the attention layer and decoder. As the decoder is not bidirectional, it initial supplied context vector is computed as:
$$h^{\prime}_0 = f(g(\forward{h_t},\backward{h_t}))$$
where $f$ is the $tanh$ activation function and $g$ is a feedforward layer.

\subsubsection{Decoder}
The decoder processes the provided hidden states $h_t$ from the encoder, and computes a decoder hidden state $h_t^{\prime}$ and the conditional distribution of the target token $y_t$ according to the following equations:

\begin{align*}
h_{t}^{\prime} &= f(h_{t-1}^{\prime}, y_{t-1}, c_{t}) \\
p(y_{t} | y_{t-1}, \ldots, y_{1}, c_{t}) &= g(h_{t}^{\prime}, y_{t-1}, c_{t})
\end{align*}
where $c_{t}$ is the context vector and $f$ and $g$ are both non-linear functions. In this research, $f$ is again a GRU like in the encoder. For the function $g$, it needs to predict probabilities between 0 and 1 and is therefore the $softmax$ function. The context vectors $c_{t}$ are computed by the Attention layer.

\subsubsection{Attention}
The attention used in this model is additive attention according to \citeauthor{bahdanau2014neural}. As the decoder section mentions, the context vectors $c_{t}$ are used in each time step $t$. The context vectors are computed as:
$$
c_{t}=\sum_{i=1}^{T} \alpha_{t i} h_{i}
$$

where the attention $\alpha$ are probabilities that are computed with the $softmax$ function:
$$
\alpha_{i j}=\frac{\exp \left(e_{i j}\right)}{\sum_{k=1}^{T} \exp \left(e_{i k}\right)} = softmax(e_{ij})
$$
The variable $e$ is the \textit{alignment model} score on how well the source ts at index $j$ matches the target at index $i$ match and is computed as:  
$$
e_{i j}=a\left(h^{\prime}_{i-1}, h_{j}\right)
$$
The function $a$ is the \textit{alignment model} and is defined as a feedforward layer in this research and is jointly trained with the total model.

\subsubsection{Embeddings}
A trainable embedding layer $E$ is used in both the encoder and decoder to get numerical representations of the source and target tokens respectively. This matrix consists of $E \in \mathbb{R}^{m \times k}$ where $m$ is the size of the corresponding vocabulary and $k$ is the embedding dimension. A token can now be looked with the one-hot encoding of the token in the vocabulary. Dropout was used after the embedding layers as this leads to a better generalization error in RNN's \cite{gal2016theoretically}.  

\subsubsection{Optimize function}
The goal of the network is to maximize the conditional log-likelihood of the target sequence given the input sequence by adjusting the model parameters:
$$
\max _{\theta} \frac{1}{N} \sum_{i=1}^{N} \log p\left(y_{i} | x_{i} ; \theta\right)
$$
where $N$ is the amount of objects in the dataset, $\theta$ are the parameters of the model, and $x_i$ $y_i$ are tokens from the source and target sequence respectively. In this research, this goal was achieved by minimizing the cross entropy loss, as this gives the same optimal set of parameters $\theta$. 

\subsection{Evaluation}
During training, the model is evaluated on the validation data by computing the cross entropy loss. The model with the lowest loss will be considered the best model. 
During testing, for all models the BLEU scores are computed according to \cite{papineni_bleu:_2001} and the ROUGE F1 scores are computed according to \cite{lin_rouge:_2004}, which is a combination of ROUGE precision and recall. The ROUGE-1 and ROUGE-2 scores are based on the overlap of unigrams and bigrams respectively. ROUGE-L is based on the longest common subsequence (LCS) and ROUGE-W builds further on this by using weighted LCSes which favors consecutive subsequences.

\subsection{Model Training}\label{subsec:training}
Extra techniques were implemented to effectively train a model that performs well. Firstly, teacher forcing is used where with a predefined probability we take the real $y_t$ instead of the predicted $\hat{y_t}$ \cite{williams1989learning}. Secondly, two techniques from NLP are applied: packing and masking. With packing, the length of the source sequence is supplied to the model such that it stops all extra padding tokens are ignored. With masking, a mask is created over all values that are not padding. This mask can then be used to in the computation for attention so that padding tokens are ignored.     

The model was trained with Stochastic Gradient Descent (SGD) with an initial learning rate of $0.1$. The learning rate was reduced with factor $0.1$ if no improvements were made on the validation loss for 10 epochs. Early stopping of the training process was done if no validation loss improvement was seen for 20 epochs. All models were trained on a Nvidia GeForce GTX 1660 GPU with 6GB of memory. After each epoch, the intermediate model was saved and the model with the least validation loss was used for evaluation.

\section{Dataset Preparation}\label{sec:preprocessing}
Sound data preprocessing is crucial for a generalizable model. In \Cref{sec:prep:existing}, the preprocessing approach from \citet{jiang_automatically_2017} is discussed. \Cref{sec:prep:new} proposes an alternative, more rigorous preprocessing technique. \Cref{sec:prep:analysis} discusses the characteristics of preprocessed datasets.

\subsection{Reference Method}\label{sec:prep:existing}

\citet{jiang_automatically_2017} use their own dataset containing 2M commits from the top 1000 Java projects on GitHub, published earlier in \cite{jiang_towards_2017}. Commit messages and diffs are cleaned and filtered, to arrive at a final dataset of 32k <commit,diff> pairs.

The dataset is filtered by removing merge and rollback commits and diffs larger than 1MB. Diffs containing more than 100 tokens and commit messages with more than 30 tokens are discarded. Then, a Verb-Direct Object filter is applied to commit messages and selects only messages starting with a verb that has a direct object dependant \cite{jiang_automatically_2017}. 

The commit messages are cleaned by extracting only the first sentence and removing issue IDs; diffs are cleaned by removing commit IDs. Both diffs and commit messages are tokenized by whitespace and punctuation \cite{jiang_automatically_2017}.



\subsection{Alternative Method}\label{sec:prep:new}
\citet{liu_neural-machine-translation-based_2018} thoroughly analyzed the dataset from \cite{jiang_automatically_2017} and found that their dataset is noisy. To improve the quality of the dataset -- the hypothesis is that more extensive preprocessing would enable a model to better learn and generalize over the relations between code changes and commit messages -- the preprocessing pipeline proposed by \cite{jiang_automatically_2017} is extended.

First, a preliminary filter is applied that removes all merge and rollback commits, which are unsuitable to be used for machine translation \cite{jiang_automatically_2017}. Then, commit messages and diffs are processed as follows.

\subsubsection{Commit Messages}

\begin{enumerate}
\item \textit{Cleaning.} GitHub issue IDs, preceding labels in the format "\textit{[Label]} Sentence.", \textit{@mentions}, URLs and SHA-1 (commit) hashes are removed from the commit messages. Furthermore, all version numbers are replaced with a placeholder token and sub-tokens (\underline{c}amel\underline{C}ase) are split. Lastly, based on \citet{liu_generating_2019}, non-English characters are removed and the commit message is lowercased.

\item \textit{Tokenizing.} Sentences is commit messages are first parsed with the NLTK Punkt sentence tokenize. Only the first sentence of the first line of a commit is retained. This sentence is then parsed by natural language toolkit SpaCy to extract tokens and their respective part-of-speech (PoS) tags. Redundant whitespace and trailing punctuation is removed.

\item \textit{Message Length Filter.} Commit messages with less than 2 or more than 30 tokens are removed.

\item \textit{Verb Filter.} Automated PoS-tagging is prone to errors if the source text uses invalid grammar. Initial experiments have shown that verbs are often classified as nouns, because developers write concise commit messages, omitting the subject of a sentence. The V-DO constraint is therefore relaxed by only requiring that a sentence starts with a verb. If the first word is not classified as verb at first, a secondary check on the message, prepended with "\textit{I}", is performed to select any remaining messages. An example construct is shown in \Cref{fig:verb_detection_error}.

\begin{figure}
    \centering
    \includegraphics[width=.75\linewidth]{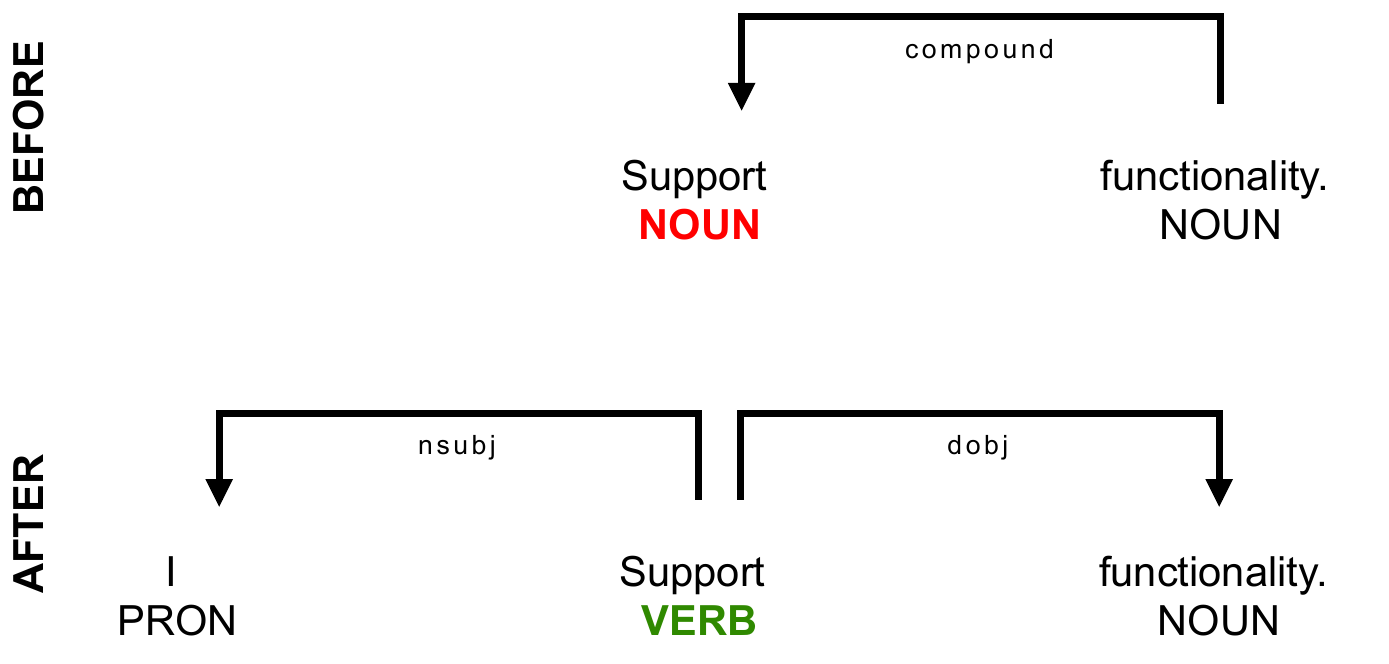}
    \caption{Automated verb detecting correction by prepending pronoun.}
    \label{fig:verb_detection_error}
\end{figure}
\end{enumerate}

\subsubsection{Diffs}
Contrary to the approach of \citet{jiang_automatically_2017}, diffs are also subject to preprocessing. The hypothesis is that diffs contain a lot of redundant information that is not informative for commit message generation.

\begin{enumerate}
    \item \textit{Parsing.} Diffs are split into blocks per file, which are processed further independently. Only files with either additions or deletions are kept.
    
    \item \textit{Cleaning.} Instead of the full path to the changed file, only the filename and extension is retained. The location of the change is removed and only the context of the change (encapsulating method or class name) is kept. Again, sub-tokens are split, non-English tokens are removed and all tokens are lowercased.
    
    \item \textit{Filtering.} Diffs with more than 100 tokens are discarded. Only changed files with a whitelisted extension are kept, changed lines from other files are removed from the diffs. Finally, the files in the diff are sorted on by most lines changed.
    
    \item \textit{Tokenizing.} Diffs are tokenized on whitespace and punctuation. The tokenizer used is an improved version of the \textit{WordPunctTokenizer}\footnote{\url{https://www.nltk.org/api/nltk.tokenize.html\#module-nltk.tokenize.punkt}}, which does not split language operators or comment indicators (e.g. \texttt{++}, \texttt{//}, etc.)
\end{enumerate}

\subsection{Processed Dataset Characteristics}\label{sec:prep:analysis}
The novel preprocessing method is applied to both the dataset collected by \citet{jiang_towards_2017} and the Java and C\# datasets collected in this work. 

\subsubsection{Dataset Sizes}
\Cref{tbl:dataset_processing} contains an overview of the processed datasets. Substantially more commits are retained from the dataset collected by \citet{jiang_towards_2017} than originally. The reason for this is twofold: \citet{jiang_automatically_2017} only keep commits starting with one of the 20 most occurring verbs, instead of applying no such filter. This makes their dataset naturally smaller. Furthermore, they discard commit messages that start with a verbs that are not classified as such by their natural language processor, instead of making an effort to better detect verbs.

\begin{table}[]
\caption{Dataset sizes before and after processing.}
\label{tbl:dataset_processing}
\begin{tabular}{@{}lll@{}}
\toprule
\textbf{Dataset} & \textbf{Original} & \textbf{Processed} \\ \midrule
         Java Top 1000    & 610K & 151K \\
         C\# Top 1000     & 1.6M & 389K \\
         NMT1 \cite{jiang_automatically_2017} & 2.1M & 32K \\
         NMT1 (processed) & 2.1M & 156K \\
         \bottomrule
\end{tabular}
\end{table}

\subsubsection{Diff Length Distribution}\label{subsec:difflength}
\citet[fig.5]{jiang_automatically_2017} have analyzed the distribution of diff lengths in their test set and found that the distribution is heavily skewed towards the maximum number of 100 tokens. \Cref{app:vis:tokendist} (a) reproduces this distribution for the convenience of the reader. Additionally, the diff length distribution is visualized (\Cref{app:vis:tokendist} (b), (c) and (d)) for every processed dataset used in this work. The distribution is remarkably different: the long tail is gone and the diff lenghts are more or less evenly distributed. This phenomenon can be explained by the removal of entire file paths but the filenames from diffs, which decreases the expected minimum length for files changed in nested folders.

\section{Results}\label{sec:results}

\subsection{Experiment parameters}
All of the models were trained with the encoder en decoder hidden dimension and embedding dimension of 512 and 256 respectively. The dropout in the embedding layers were set to $0.1$. The input and output dimension of the model differ per dataset as it contains a different amount of unique tokens based on the generated vocabulary. The vocabulary sizes of each dataset are shown in \Cref{tbl:vocabs} and these correspond to the input and output dimensions of the trained model. The batch size was 64 for all models.

Note that the hidden and embedding dimensions of the model are half of the model that was trained by \citeauthor{jiang_automatically_2017}. This was due to the memory limitations of the GPU that was used.
\begin{table}[]
\caption{The vocabulary parameters for each dataset, where the number indicate the amount of unique tokens in the source or target vocabulary.}
\label{tbl:vocabs}
\begin{tabular}{@{}lll@{}}
\toprule
\textbf{Dataset}                   & \textbf{Source} & \textbf{Target} \\ \midrule
Java Top 1000                      & 30,854                      & 13,871                      \\
C\# Top 1000                       & 24,251                      & 11,382                      \\
NMT1 \cite{jiang_automatically_2017} & 50,004                      & 14,200                      \\
NMT1 - processed                   & 28,672                      & 14,817                      \\ \bottomrule
\end{tabular}
\end{table}

\subsection{Testing performance}
After training the model with the settings mentioned above, and according to the training procedure in \Cref{subsec:training}, the model was evaluated on the testing set. Both the BLEU and ROUGE score were computed and are shown in \Cref{tbl:results}. It can be seen that the testing results on the dataset that were collected and preprocessed in this research are significantly lower. The results on the dataset from \citeauthor{jiang_automatically_2017} are somewhat similar: 33.63 against the reported 31.92 in \cite{jiang_automatically_2017} 

\begin{table*}[]
\caption{The evaluation results on the testing dataset for each dataset.}
\label{tbl:results}
\begin{tabular}{@{}llllll@{}}
\toprule
\textbf{Dataset} & \textbf{BLEU} & \textbf{ROUGE-1} & \textbf{ROUGE-2} & \textbf{ROUGE-L} & \textbf{ROUGE-W} \\ \midrule
Java top 1000                        & 5.33  & 23.60   & 10.87   & 26.52   & 19.35   \\
C\# top 1000                         & 7.31  & 26.84   & 13.16   & 29.85   & 22.08   \\
NMT1 \cite{jiang_automatically_2017} & \textbf{33.63} & \textbf{37.20}   & \textbf{23.22}   & \textbf{40.01}   & \textbf{30.10}   \\
NMT1 - processed                     & 3.19  & 20.26   & 7.93   & 23.05   & 16.37   \\ \bottomrule
\end{tabular}
\end{table*}

\section{Discussion}\label{sec:discussion}
The results in \Cref{tbl:results} show a significant difference between the obtained performance between the NMT1 dataset from \cite{jiang_automatically_2017} and any other dataset. In this section a qualitative analysis is done to explain these results. Also, discussion points of this research and ideas for future research will be mentioned.

\subsection{Qualitative analysis NMT1}
One pattern in the commit message that is commonly seen in the testing dataset of NMT1 is \texttt{ignore update '<filename>'}, where \texttt{<filename>} differs among commits. A total of $12.76\%$ in the testing set has this pattern and the model was able to classify them all correctly. A visualisation of the attention of one of these examples is shown in \Cref{app:vis:ignore}. It can be seen that the model is able to attend to the path names at the start of the input sequence and copy the path name to the output to produce the commit message correctly. When we look at another message \texttt{prepared version 0 . 2 - snapshot} in \Cref{app:vis:prepare}, for which the prediction was \texttt{prepare next development version.}, the attention is all focused on one token in the input namely the slash token. The model is unable to generate the correct output tokens from input tokens.

Still our model achieved a better BLEU score of 33.63 compared to the 31.92 in \cite{jiang_automatically_2017}. The only difference that was made in this research was to lower case all of the input tokens during vocabulary generation. This led to a reduced output dimension of 14200 compared to the 17000 in \cite{jiang_automatically_2017}. Thus the problem was less computational expensive and our model was able to achieve better results.

When the preprocessing from \Cref{sec:preprocessing} is applied on the NMT1 dataset, only the filename and his extension are retrained from the full path name. This means that the model can not simply copy the input tokens to the output tokens anymore. This degrades the model performance as a high percentage of the testing set was in this pattern.

On both the Java and C\# dataset that was collected in this research, the trained model also performs rather poorly. Regardless of the programming language that the model is trained upon, the performance is significantly worse than achieved on the dataset from \cite{jiang_automatically_2017}. We conclude that this performance difference comes from the fact that the model tries to learn to translate long diff sequence into short message sequences, something it is unable to do. The testing dataset from \cite{jiang_automatically_2017} contains many easier examples than a real world dataset collected from GitHub. It is still unclear to us how \citeauthor{jiang_automatically_2017} created the training testing split for their dataset and this likely has a high influence.

\subsection{Discussion points}
Certain points in this research are subjected to some critical discussion. Firstly, the models that were trained in this research had a lower dimensionality than in \cite{jiang_automatically_2017} due to GPU limitations. It is expected that the same kind of results will be achieved if these dimensions are higher, as the model tries to solve a problem that is unsolvable. 

Another fact is that during translation to a prediction, the tokens are generated in a greedy fashion and the token with the highest probability is selected. Another approach to do this would be beam-search, in which multiple option sets are explored to find the set that has the highest likelihood. This could lead to better translations.

\subsection{Future research}
One of the problems of this research is that a sequence of tokens in the form of a \texttt{git diff} file is unable to capture the structure of the code changes. An interesting approach to this problem would be to embed the code before and after the code changes, and subtract or concatenated these embeddings to have a vector representation of the code changes. However, this would require a code embedding that can embed multiple functions or files into a single vector that retains the information. More research in embedding the code properly could lead to interesting results and message generation.

Another point to improve upon in future research could be to first classify commits into multiple categories such as additions, deletions, and refactors. It is hypothesized that these commits have a structural difference among them, and training different models could lead to exploitation of these factors and hopefully to better results.

\section{Conclusion}\label{sec:conclusion}
The purposes of the current research were (1) to determine if the neural approach to generate commit messages from code changes, as presented by \citet{jiang_automatically_2017}, was reproducible and (2) to investigate if more rigorous preprocessing techniques would improve the performance of the model.

Experiments showed that a reproduction of the attentional RNN encoder-decoder model from \citet{jiang_automatically_2017} achieves slightly better results on the same dataset. This confirms the reproducibility of \cite{jiang_automatically_2017} under similar circumstances.

To answer the second question, an alternative preprocessing method was proposed in an effort to better clean and remove noisy commits from the original dataset. Furthermore, two new datasets were collected from GitHub, one containing commits from the Top 1000 Java projects and one with commits from the Top 1000 C\# projects, to compare the impact of the novel preprocessing on different datasets.

However, the model was unable to generate commit messages of high quality for any input dataset that was processed with the novel technique. The BLEU score dropped by at least 78\% for any dataset. This exposed the underlying problem of the original model, which seems to score high by remembering (long) path names and frequently occurring messages from the training set.

Automated commit message generation is therefore still very much an open problem. Different code change embeddings, for example by embedding the before and after state of the code separately, or focusing on specific types of commits, could improve the quality of generated commit messages in the future.



\section{Reflection}
Before arriving at our current approach, we had some other ideas about how we could tackle this problem. We looked at existing models,  such as Word2Vec \cite{mikolov2013efficient}, Code2Vec \cite{alon2018code2vec} and Code2Seq \cite{alon2018code2seq}. The idea was to use these models to embed the code before and after a commit and use a combination of these embeddings to represent the change in the code. Then, we could train a model on this embedding of the change to generate commit messages. 

In the end, it was not feasible to implement this for a set of (partial) code changes, of which a diff consists. This would result in a variable amount of change embeddings, which would be hard to combine into a single embedding which would still represent the commit. Also, while experimenting with Code2Vec and Code2Seq, we encountered the limitation of only being able to embed small functions and no full source code files. This made both models unusable for our problem.

With regard to training models, we had to make some compromises. We lowered the amount of dimensions for our reproduction of the model of \citeauthor{jiang_automatically_2017} \cite{jiang_automatically_2017}, because of memory limitations. Since we could only train on one PC -- with one GPU -- that was powerful enough, we did not have time to train all the models that would have made an interesting comparison. An improvement for future editions of this course could be to provide credits for cloud services, which can potentially be acquired for free for academic purposes.

Also, two weeks before the deadline, one of our team members unfortunately had to leave the team, which left us with more work to do than we expected.

\bibliographystyle{ACM-Reference-Format}
\bibliography{bibliography}

\onecolumn\appendix

\section{Visualized Attention}

\subsection{Ignore update pattern from NMT1}\label{app:vis:ignore}
\begin{itemize}
    \item \textbf{True message:} ignore update ' modules / apps / foundation / login / .
    \item \textbf{Predicted message:} ignore update ' modules / apps / foundation / login / .
\end{itemize}

\begin{figure}[H]
    \centering
        \includegraphics[width=\textwidth]{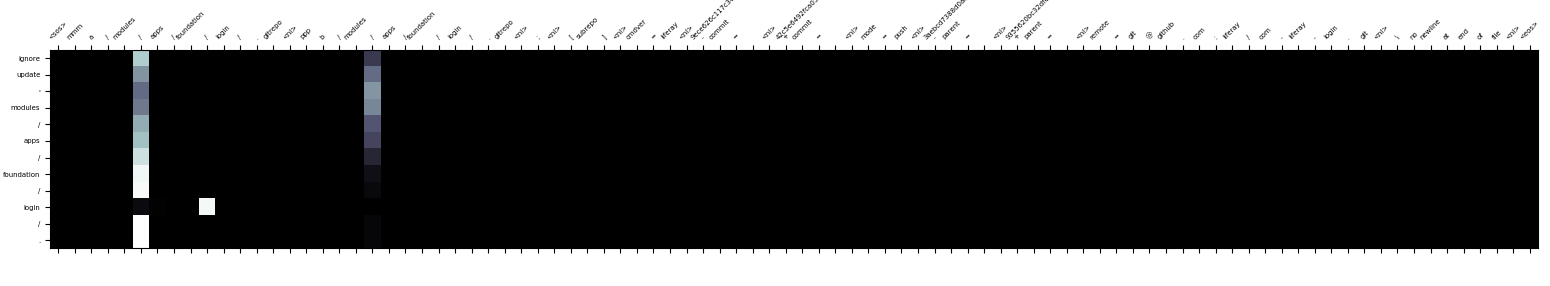}
        \caption{Attention visualised for a sentence that has the \texttt{ignore update '<filename>'} pattern. The model is able to attend to the specific words in the path name in the diff file to generate the correct label.}
        \label{fig:ignore}
\end{figure}
    
\subsection{Another message from NMT1}\label{app:vis:prepare}
\begin{itemize}
    \item \textbf{True message:} prepared version 0 . 2 - snapshot .
    \item \textbf{Predicted message:} prepare next development version .
\end{itemize}
\begin{figure}[H]
        \includegraphics[width=\textwidth]{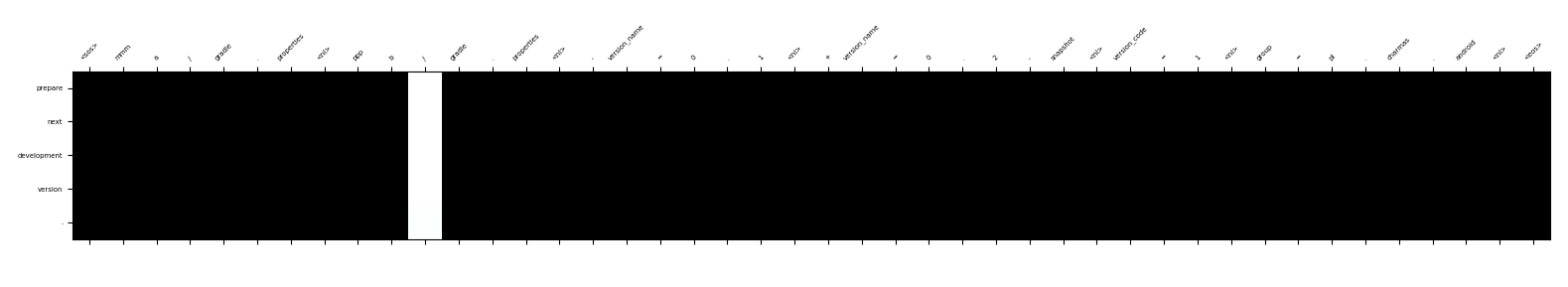}
        \caption{Attention visualised for a selected example in the testing set. Although the predicted message is close to the real message, the model attends to random parts of the input sequence.}
        \label{fig:prepare}
\end{figure}

\subsection{Distribution of amount of tokens in diffs in the test sets}
\begin{figure}[H]

        \includegraphics[width=.45\textwidth]{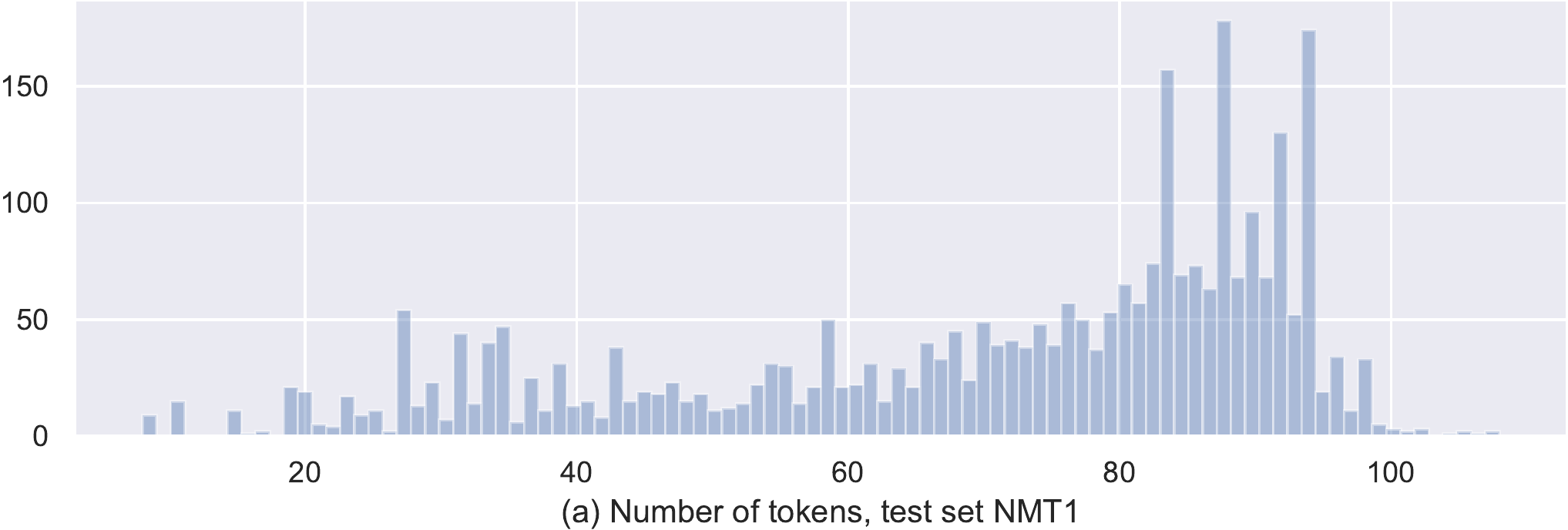}\hfill
        \includegraphics[width=.45\textwidth]{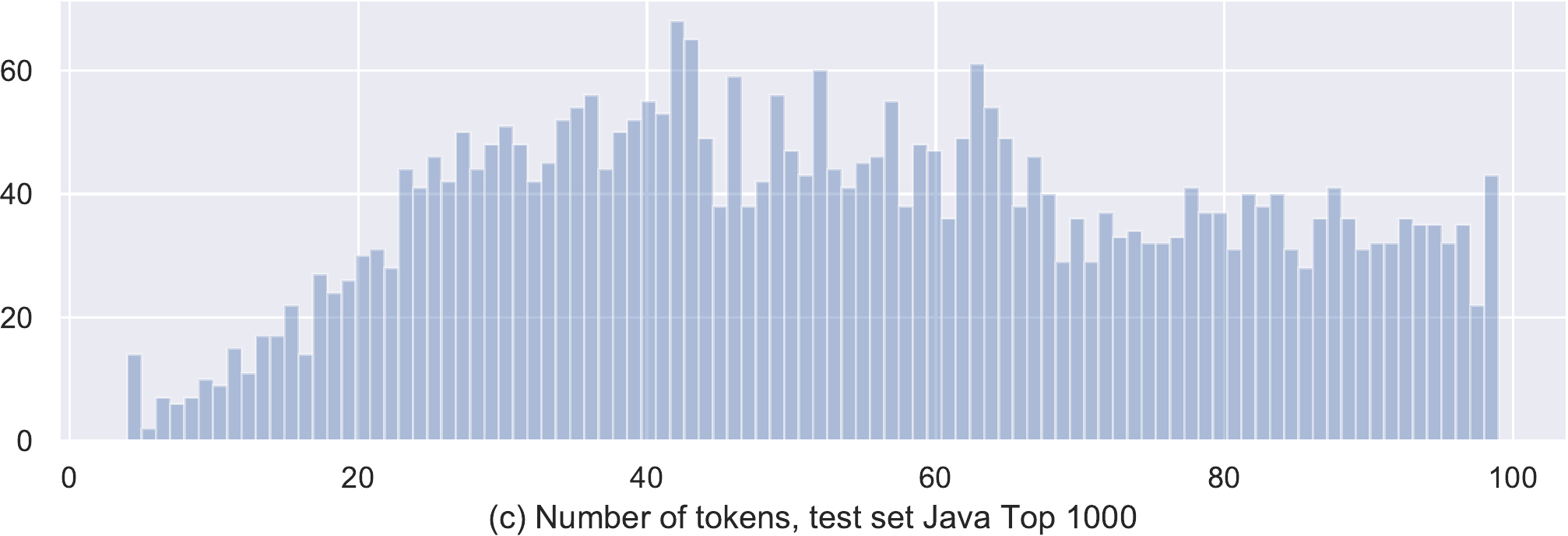} \\[.5cm]
    \includegraphics[width=.45\textwidth]{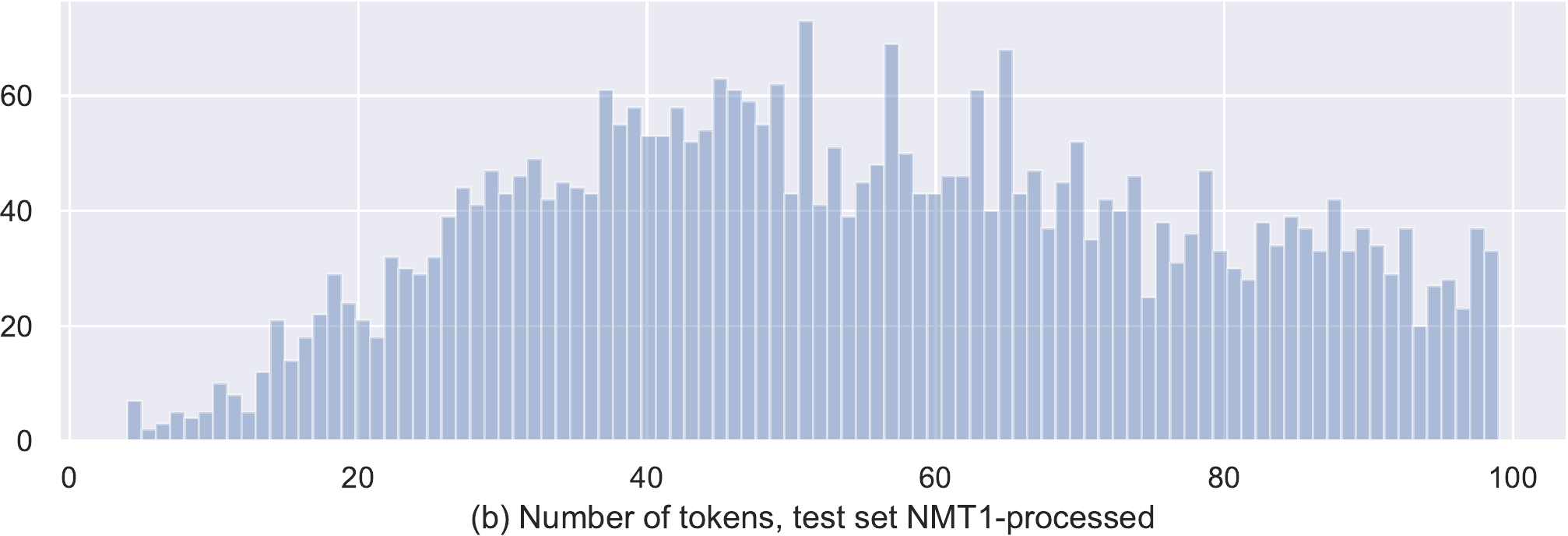}\hfill
    \includegraphics[width=.45\textwidth]{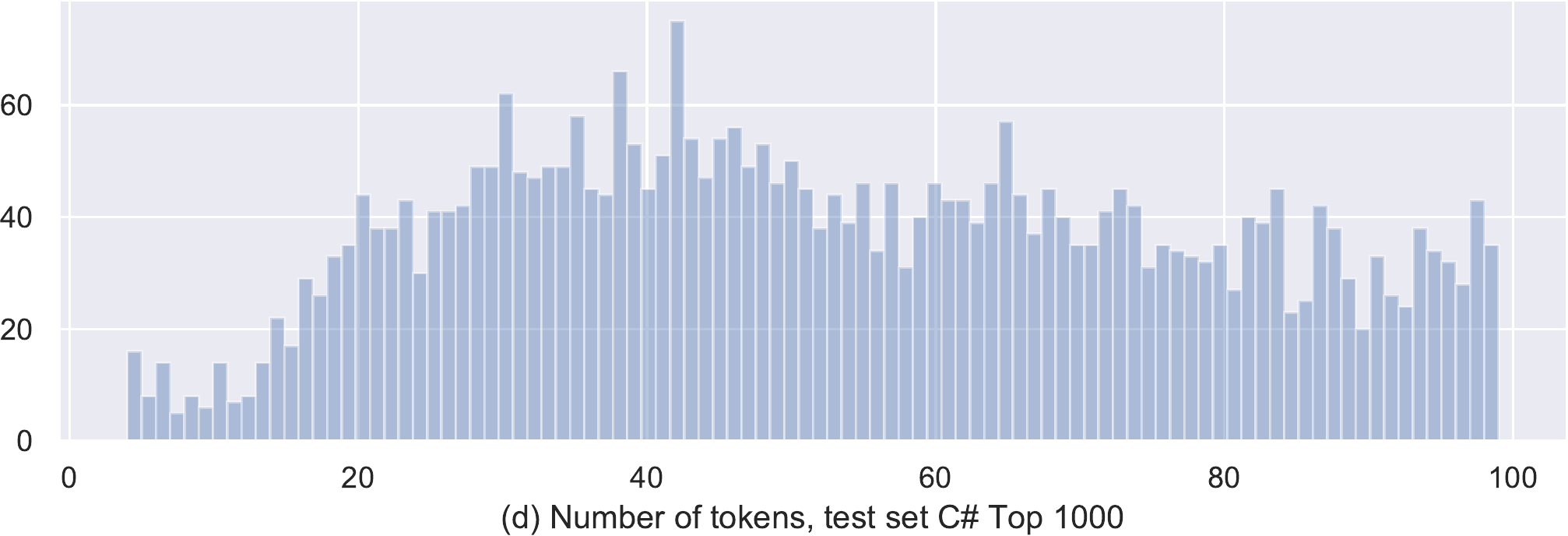}
    \caption{Distribution of amount of tokens in diffs in the test sets}\label{app:vis:tokendist}
\end{figure}

\end{document}